\documentclass[12pt]{iopart}

\usepackage{iopams}
\usepackage{graphics}

\newcommand{\im}[0]{{\rm i}}
\newcommand{\iint}[0]{\int\!\!\!\int}
\newcommand{\iiint}[0]{\int\!\!\!\int\!\!\!\int}
\newcommand{\iiiint}[0]{\int\!\!\!\int\!\!\!\int\!\!\!\int}

\begin{document}

\title[Lindblad equation for scintillation]{Lindblad equation for the decay of entanglement due to atmospheric scintillation}
\author{Filippus S. Roux}
\address{CSIR National Laser Centre, P.O. Box 395, Pretoria 0001, South Africa}
\ead{fsroux@csir.co.za}
\begin{abstract}
The quantum state for the spatial degrees of freedom of photons propagating through turbulence is analyzed. The turbulent medium is modeled by a single phase screen for weak scintillation conditions and by multiple phase screens for general scintillation conditions. In the former case the process is represented by an operator product expansion, leading to an integral expression that is consistent with current models. In the latter case the evolution of the density operator is described by a first order differential equation with respect to the propagation distance. It is shown that this differential equation has the form of a Lindblad master equation. Additionally, it is shown that this differential equation can take on the form of the infinitesimal propagation equation.
\end{abstract}


\submitto{\JPA}


\section{Introduction}

Studying how photonic states that are initially entangled in terms of their transverse spatial modes loose their entanglement as they propagate through atmospheric turbulence, one finds oneself within the overlap between two fields. One is the scintillation of classical optical beams, which has been studied for several decades \cite{tatarski0,tatarskii,aobook1,scintbook}. The other is the currently vibrant field of quantum information science \cite{nc} and in particular the study of open quantum systems \cite{petru} and quantum optics \cite{mw}. 

The problem of entanglement decay due to atmospheric scintillation has been considered by Paterson \cite{paterson} from a conventional classical optics point of view, by assuming that one can represent the turbulent atmosphere as a single phase screen. This single phase screen model is currently used as the basis for most work that is being done on the evolution of entangled photons propagating through atmospheric turbulence \cite{sr,qturb4,qturb3}. However, it is by construction only valid under weak scintillation conditions. Although the predictions of this approach have been shown to be consistent with experimental observations \cite{oamturb}, one can rightfully ask whether this approach makes sense within the context of quantum information theory:\ does it represent a valid quantum process?  More recently, a multiple phase screen model has been proposed \cite{ipe,iperr}. However, it is also derived using a classical optics approach and as such its validity as a quantum process also comes under question. In both these cases the density matrix that is obtained as solution represents a physical quantum state only if it has a unity trace and no negative eigenvalues.

In this paper we consider the evolution of the photonic quantum states propagating through turbulence from the perspective of quantum information theory. The aim is to develop a formalism with which one can study the decay of the transverse spatial modal entanglement of photonic states. Within this context we consider both the single phase screen approach, which assumes weak scintillation, and the multiple phase screen approach, without the assumption of weak scintillation. The single phase screen process, which one can express as an operator product expansion, leads to the same integral expression found from the classical optics derivation \cite{paterson}. In the case of the multiple phase screen process, the resulting expression is in the form of a Lindblad equation \cite{lindblad0,lindblad,nc,petru}. The latter indicates that the formalism represents a quantum process that produces a valid density operator. We also show that this result reproduces the equation in \cite{ipe}. However, the aim here is not to rederive the infinitesimal propagation equation (IPE) obtained in \cite{ipe}, but to obtain an equation that represents a valid quantum process, as given by the Lindblad equation.

The Laguerre-Gaussian modes, with their discrete radial and azimuthal indices, are a popular choice for the spatial modes used in free-space quantum communication systems. However, it has been argued \cite{boyd} that it may be more beneficial to use a plane wave basis. The analysis that is presented here does not assume any particular set of transverse spatial modes and can be employed for any such modal basis.

The main part of the derivations is done for a single photon. Since a single photon cannot represent an entangled quantum state, the same derivation would be valid for a classical optical field. However, by using a quantum mechanics formalism, one can readily generalized the single photon expressions to expressions for multiple photon quantum state, as demonstrated in Sec.~\ref{genipe}, which then allows quantum entanglement.

The paper is organized as follows. First some basic aspects of turbulence and scintillation are reviewed in Sec.~\ref{agter}. In Sec.~\ref{pmod} we consider the single phase screen model from the perspective of a quantum process and in Sec.~\ref{weak} we discuss the conditions for weak scintillation. In Sec.~\ref{ipesec} we derive the Lindblad equation using the multiple phase screen approach. Finally, we provide some conclusions in Sec.~\ref{concl}.

\section{Turbulence and scintillation}
\label{agter}

The effect of atmospheric turbulence on an optical beam propagating through it, is introduced by fluctuations in the refractive index of the air. The refractive index of a turbulent atmosphere can be expressed as  $n=1+\tilde{n}({\bf r})$, where the average refractive index is approximately equal to $1$ and the small fluctuations are represented by $\tilde{n}({\bf r})$. 

Scintillation is the process of distortion that an optical beam experiences while propagating through a random medium, such as a turbulent atmosphere.  A quantitative model for optical scintillation requires a model for the turbulence. There are a number of such models \cite{tatarski0,tatarskii,aobook1,scintbook} depending, among other things, on whether one includes the effects of the inner scale and the outer scale. (The inner and outer scales introduce cut-offs at high and low spatial frequencies, respectively.) The simplest model is the Kolmogorov model, which is valid in the inertial range between the inner and outer scales. The Kolmogorov power spectral density of the refractive index fluctuations is given by \cite{scintbook,iperr}
\begin{equation}
\Phi_{\rm n} ({\bf k}) = 0.033 (2\pi)^3 C_{\rm n}^2 |{\bf k}|^{-11/3} ,
\label{klmgrv}
\end{equation}
where $C_{\rm n}^2$ is the refractive index structure constant. The strength of the turbulence is determined by $C_{\rm n}^2$, with values ranging from about $10^{-17}$~m$^{-3/2}$ for weak turbulence to about $10^{-13}$~m$^{-3/2}$ for strong turbulence. The Wiener-Khinchine theorem
\begin{equation}
\Phi_{\rm n} ({\bf k}) = \int B({\bf r}) \exp( -\im {\bf k}\cdot{\bf r})\ {\rm d}^3 r ,
\label{wiener}
\end{equation}
gives the relationship between the power spectral density and the autocorrelation function $B(\Delta{\bf r}) = {\cal E}\{ \tilde{n}({\bf r}_1) \tilde{n}({\bf r}_2) \}$, where $\Delta{\bf r}={\bf r}_1-{\bf r}_2$ and ${\cal E}\{ \cdot \}$ is the ensemble average. The structure function, which is related to the autocorrelation function, is defined by 
\begin{equation}
D(\Delta{\bf r}) = {\cal E}\left\{ \left[ \tilde{n}({\bf r}_1) - \tilde{n}({\bf r}_2) \right]^2 \right\} = 2\left[B(0)-B(\Delta{\bf r})\right] ,
\label{structf}
\end{equation}
and can be obtained from interference measurements.

The strength of scintillation is not only determined by the strength of the turbulence, but also by the other relevant dimension parameters, namely, the distance $z$ over which the light propagates through turbulence and the wavelength of the light $\lambda$. These parameters are combined into the Rytov variance, given by
\begin{equation}
\sigma_R^2 = 1.23 C_{\rm n}^2 k^{7/6} z^{11/6} ,
\label{rytov}
\end{equation}
where $k$ is the wave number ($2\pi/\lambda$). For plane waves, strong scintillation conditions is believed to exist when $\sigma_R^2>1$ and, for Gaussian beams (with radius $\omega_0$), strong scintillation exists when \cite{scintbook}
\begin{equation}
\sigma_R^2 > \left(t+{1\over t}\right)^{5/6} ,
\label{sterksint}
\end{equation}
where $t = z/z_R$, with $z_R$ being the Rayleigh range ($\pi\omega_0^2/\lambda$).

\section{Single phase screen model}
\label{pmod}

Although it is not directly formulated in the language of quantum information theory, the single phase screen model \cite{paterson} represents a valid quantum process --- it can be expressed as an operator product expansion \cite{nc}. Under weak scintillation conditions, the turbulent atmosphere can be represented by a single phase modulation \cite{paterson}. The corresponding quantum process is a single step process 
\begin{equation}
\rho(z) = U \rho(0) U^{\dag} , 
\label{quop}
\end{equation}
where the unitary operator is represented by a single phase factor $U \sim \exp[\im\theta(x,y)]$. 

Assuming that the original input density matrix is a pure state $\rho(0)=|\psi\rangle \langle\psi |$, one can express the individual output density matrix elements by
\begin{equation}
\rho_{mn}(z) = \langle m|U|\psi\rangle \langle\psi |U^{\dag}|n\rangle ,
\label{rho0}
\end{equation}
where $|m\rangle$ represents a discrete orthogonal basis for the transverse modes, such as the Laguerre-Gaussian modes. We insert an identity operator, resolved in terms of the two-dimensional spatial coordinate basis, to get
\begin{eqnarray}
\langle m|U|\psi\rangle & = & \int \langle m|{\bf r}\rangle \langle{\bf r}|U|\psi\rangle\ {\rm d}^2 r \label{oper} \\
\langle\psi |U^{\dag}|n\rangle & = & \int \langle\psi |U^{\dag}|{\bf r}\rangle \langle{\bf r}|n\rangle\ {\rm d}^2 r , \label{operc}
\end{eqnarray}
where ${\bf r}$ is the two-dimensional transverse position vector. The mode functions for the transverse spatial modes are given by $E_n({\bf r}) = \langle{\bf r}|n\rangle$ and the single phase screen approximation leads to $\langle{\bf r}|U|\psi\rangle = \exp[\im\theta({\bf r})] \psi({\bf r})$, where $\psi({\bf r})=\langle{\bf r}|\psi\rangle$ is the input field and $\theta({\bf r})$ is the phase function that is obtained from the refractive index fluctuations $\tilde{n}$ by an integration along the direction of propagation
\begin{equation}
\theta(x,y) = k \int_0^z \tilde{n}(x,y,z)\ dz .
\label{faseint}
\end{equation}
The magnitude of the random phase modulation is therefore proportional to the propagation distance $z$.

The expression for the density matrix element in Eq.~(\ref{rho0}) then becomes
\begin{equation}
\fl \rho_{mn}(z) = \iint E_m^*({\bf r}_1) E_n({\bf r}_2) \psi({\bf r}_1) \psi^*({\bf r}_2) \exp\left[\im\theta({\bf r}_1)-\im\theta({\bf r}_2)\right]\ {\rm d}^2 r_1\ {\rm d}^2 r_2 .
\label{rho2}
\end{equation}

The unitary operator $U$ in Eq.~(\ref{quop}) corresponds to a particular instance of the refractive index fluctuations. Since we do not presume to have complete knowledge of the medium at any particular time and therefore can only make statistical predictions about its effect, we need to compute the ensemble average over all refractive index fluctuations. The density matrix elements are therefore given by 
\begin{equation}
{\cal E}\left\{\rho(z)\right\} = {\cal E}\left\{ U \rho(0) U^{\dag} \right\} = \sum_s^S {1\over S}\ U_s \rho(0) U_s^{\dag} , 
\label{quopsom}
\end{equation}
where we show the explicit summation done to evaluate the ensemble average, with the subscript $s$ denoting a particular element of the ensemble (a particular instance of the refractive index fluctuations) and $S$ being the total number of elements in the ensemble. Note that Eq.~(\ref{quopsom}) represents an operator product expansion. When the ensemble average is applied to the expression in Eq.~(\ref{rho2}), it only affects the exponential function containing the random phase modulations and leads to 
\begin{equation}
{\cal E}\left\{ \exp\left[\im\theta({\bf r}_1)-\im\theta({\bf r}_2)\right] \right\} = \exp\left[-{1\over 2} D \left(|{\bf r}_1-{\bf r}_2|\right) \right] .
\label{expavg}
\end{equation}
Here $D(\cdot)$ is the phase structure function. For Kolmogorov turbulence, it is given by 
\begin{equation}
D(x) = 6.88 \left({x\over r_0}\right)^{5/3} ,
\label{strukt}
\end{equation}
in terms of the Fried parameter \cite{fried},
\begin{equation}
r_0 = 0.185 \left( {\lambda^2\over C_{\rm n}^2 z}\right)^{3/5} .
\label{fried}
\end{equation}

The ensemble average of the density matrix element is therefore given by 
\begin{equation}
\fl {\cal E}\left\{ \rho_{mn}(z) \right\} = \iint E_m^*({\bf r}_1) E_n({\bf r}_2) \psi({\bf r}_1) \psi^*({\bf r}_1) \exp\left[-{1\over 2} D \left( |\Delta{\bf r}| \right) \right]\ {\rm d}^2 r_1\ {\rm d}^2 r_2 .
\label{rho3}
\end{equation}
One can render the integration variables in Eq.~(\ref{rho3}) dimensionless by normalizing ${\bf r}_1$ and ${\bf r}_2$ with respect to some dimension parameter, such as the beam radius $\omega_0$. The resulting expression that follows from the integral thus only depends on the dimensionless combination $\omega_0/r_0$. All the adjustable dimension parameters are contained in $\omega_0/r_0$, including the propagation distance $z$. As a result, the complete $z$-dependence is determined by the $\omega_0/r_0$-dependence inside the structure function.

The integral in Eq.~(\ref{rho3}) is not analytically tractable due to the power of $5/3$ that appears in the structure function inside the argument of the exponential function. Often this problem is avoided by approximating the structure function in Eq.~(\ref{strukt}) by a quadratic structure function
\begin{equation}
D \sim \left({x\over r_0}\right)^{5/3} \rightarrow \left({x\over r_0}\right)^2 .
\label{kwadstruk}
\end{equation}

Smith and Raymer \cite{sr} used the single phase screen model with the quadratic structure function approximation to calculate the concurrence \cite{wootters2} (a measure of qubit entanglement) as a function of $\omega_0/r_0$ for a photon pair that is initially entangled as a Bell state in terms of the azimuthal index of the Laguerre-Gauss modes. They found that the concurrence decays to zero when $\omega_0/r_0 \approx 1$. For larger values of the azimuthal index the concurrence survives up to larger values of $\omega_0/r_0$. This has been confirmed by experimental measurements and numerical simulations \cite{oamturb}.

\section{Weak scintillation}
\label{weak}

\begin{figure}[ht]
\includegraphics{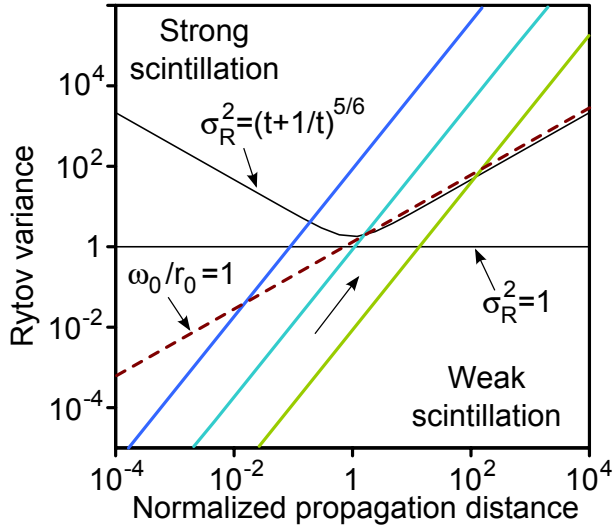}
\caption{The different regions of scintillation strength are shown in terms of the Rytov variance $\sigma_R^2$ as a function of the normalized propagation distance $t$ on a log-log scale. The boundaries between the region of weak scintillation at the bottom and the region of strong scintillation at the top, are shown for both plane waves ($\sigma_R^2=1$) and Gaussian beams [$\sigma_R^2=(t+1/t)^{5/6}$]. The dashed line represents the line where $\omega_0/r_0=1$. The three slanted coloured lines represent the scintillation strengths of optical beams propagating through different strengths of turbulence --- from left to right:\ $C_{\rm n}^2=\{10^{-12}, 10^{-14}, 10^{-16}\}$ m$^{2/3}$.}
\label{rytovfig}
\end{figure}

Let's consider more carefully the conditions under which one can use the single phase screen approximation. It is stated that the single phase screen approximation assumes that the scintillation is weak. The strength of the scintillation is quantified by the Rytov variance $\sigma_R^2$, defined in Eq.~(\ref{rytov}). In Fig.~\ref{rytovfig}, a diagram is used to represent the different regions in terms of the Rytov variance as a function of the normalized propagation distance $t$. Both axes are shown in logarithmic scales. For an optical beam propagating through turbulence of a particular strength, the Rytov variance is proportional to $t^{11/6}$. In Fig.~\ref{rytovfig} three lines are shown that represent the Rytov variance for optical beams propagating through turbulence with three different strengths. These lines start at the bottom of the diagram in the region of weak scintillation and move toward the region of strong scintillation at the top of the diagram. These lines cross the boundary between these regions, given by Eq.~(\ref{sterksint}), more or less at the same point where they cross the line where $\omega_0/r_0=1$ (dashed line in Fig.~\ref{rytovfig}). The dashed line is obtained from the Rytov variance by expressing it in terms of $\omega_0/r_0$,
\begin{equation}
\sigma_R^2 = 1.637\ t^{5/6} \left({\omega_0\over r_0}\right)^{5/3} .
\label{rytov1}
\end{equation}
Hence, when $\omega_0/r_0=1$ the Rytov variance is proportional to $t^{5/6}$.

According to \cite{sr}, the region where $\omega_0/r_0=1$ is approximately where the concurrence goes to zero. Therefore, the optical field never seems to arrive in the region of strong scintillation with nonzero entanglement, which seems to imply that the single phase screen model can be used for all situations. There are, however, some conditions under which the entanglement can survive into the region of strong scintillation, for instance, when larger values of azimuthal index are used \cite{sr}, or when the states are optimized to retain their entanglement over larger distances \cite{bruenner}. As a result, it does make sense to develop a reliable model that does not only apply in weak scintillation conditions.

\section{Multiple phase screen approach}
\label{ipesec}

\subsection{Operator product expansion}

In the multiple phase screen approach the quantum process is broken up into multiple steps. Instead of doing the calculation in one step, going from the pure initial state to the final mixed state, as was done in Sec.~\ref{pmod}, one can break the process up into infinitesimally small steps. In each step the infinitesimal propagation process operates on the density operator of a (potentially) mixed state and produces a slightly perturbed version of this density operator
\begin{equation}
\rho(z) \rightarrow \rho(z+dz) = dU \rho(z) dU^{\dag} .
\label{infquop}
\end{equation}
Here $dU$ is the unitary operator for an infinitesimal propagation through turbulence. The ensemble averaging can be expression as a summation over the ensemble elements
\begin{equation}
 {\cal E}\left\{\rho(z+dz)\right\} = {\cal E}\left\{ dU \rho(z) dU^{\dag} \right\} = \sum_s^S {1\over S}\ dU_s \rho(z) dU_s^{\dag} ,
\label{infquopsom}
\end{equation}
where, as before, the subscript $s$ denotes a particular instance of the refractive index fluctuations. In terms of the density matrix elements for a single photon, this becomes
\begin{equation}
\rho_{mn}(z+dz) = \sum_s^S {1\over S}\ \sum_{pq} \langle m|dU_s|p\rangle \rho_{pq}(z) \langle q|dU_s^{\dag}|n\rangle .
\label{infqelsom}
\end{equation}
We'll first consider the density matrix for a single photon and then generalize it to the case with two photons.

\subsection{Field transformation}

Although the single phase screen approximation is valid for each step in this situation, thanks to the short propagation distances, the fact that this operation is performed repeatedly implies that one also needs to include the free space propagation process, which is neglected in the single phase screen model.\footnote{One can neglect the propagation process in the single phase screen model, because the inner products before and after a single propagation will give the same result.} As a result, $dU$ will contain a propagation term, in addition to the random phase modulation term. To find the correct expression for $dU$, we use the equation of motion in turbulence, given by \cite{scintbook}
\begin{equation}
\nabla_T^2 g({\bf x}) - \im 2k\partial_z g({\bf x}) + 2k^2 \tilde{n}({\bf x}) g({\bf x}) = 0 .
\label{eomturb}
\end{equation}
Here $g({\bf x})$ represents the scalar electromagnetic field and $\tilde{n}({\bf x})$ is the fluctuations in the refractive index. The expression in Eq.~(\ref{eomturb}) is obtained from the Helmholtz equation under the paraxial approximation, which assumes that the optical field propagates close to the beam axis (in this case the $z$-axis) and under the assumption that $\tilde{n} \ll \langle n \rangle \approx 1$. The conditions for both these approximations are well satisfied in our situation.

In the two-dimensional transverse Fourier domain, Eq.~(\ref{eomturb}) becomes 
\begin{equation}
-4\pi^2|{\bf a}|^2 G({\bf a},z) - \im 2 k \partial_z G({\bf a},z) + 2k^2 N({\bf a},z) \star G({\bf a},z) = 0 ,
\label{fteomturb}
\end{equation}
where ${\bf a}$ is the two-dimensional spatial frequency vector (related to the two-dimensional momentum vector by ${\bf K}=2\pi{\bf a}$), $\star$ denotes convolution and $G({\bf a},z)$ and $N({\bf a},z)$ are the two-dimensional transverse Fourier transforms of $g({\bf x})$ and $\tilde{n}({\bf x})$, given by
\begin{equation}
G({\bf a},z) = \int g({\bf x}) \exp[\im 2\pi(ax+by)]\ {\rm d} x {\rm d} y
\label{ftg}
\end{equation}
and
\begin{equation}
N({\bf a},z) = \int \tilde{n}({\bf x}) \exp[\im 2\pi(ax+by)]\ {\rm d} x {\rm d} y ,
\label{ftn}
\end{equation}
respectively. It then follows that
\begin{equation}
\fl G({\bf a},z_0+dz) = G({\bf a},z_0) + { \im dz \over 2k} \left[ 4\pi^2 |{\bf a}|^2 G({\bf a},z_0) - 2 k^2 N({\bf a},z0) \star G({\bf a},z_0) \right] .
\label{transgft}
\end{equation}
One can use $G({\bf a},z)$ to define a quantum state in terms of a two-dimensional momentum basis. For instance, for some basis state $|m\rangle$ we have
\begin{equation}
|m\rangle = \int |{\bf a}\rangle G_m({\bf a},z)\ {{\rm d}^2 a} ,
\label{modef}
\end{equation}
where $|{\bf a}\rangle$ ($\equiv |{\bf K}\rangle$) denotes the two-dimensional momentum basis elements and $G_m({\bf a},z) = \langle{\bf a}|m\rangle$. Note that, if we substitute $G({\bf a},z_0)=G_m({\bf a},z_0)$ in Eq.~(\ref{transgft}), then one cannot assume that the transformed wave function is still associated with the same basis element $G({\bf a},z_0+dz) \neq G_m({\bf a},z_0+dz)$. This is due to the distortion introduced by the noise term that contains $N({\bf a},z_0)$.

Finally, we use Eqs.~(\ref{transgft}) and (\ref{modef}) to express the quantum operation for the infinitesimal propagation through turbulence as
\begin{eqnarray}
\langle m | dU_s | p \rangle & = \delta_{mp} + {\im dz \over 2k} \int G_m^*({\bf a},z_0) \left[4\pi^2|{\bf a}|^2 G_p({\bf a},z_0) \right. \nonumber \\
& \ \ \left. - 2 k^2 N_s({\bf a},z_0) \star G_p({\bf a},z_0) \right]\ {{\rm d}^2 a} \nonumber \\
& = \delta_{mp} + \im dz\ {\cal P}_{mp} + dz\ {\cal L}_{s,mp} ,
\label{infudef}
\end{eqnarray}
where, in the last line, we defined
\begin{equation}
{\cal P}_{mp}(z) = {2\pi^2 \over k} \int |{\bf a}|^2 G_m^*({\bf a},z) G_p({\bf a},z)\ {{\rm d}^2 a}
\label{kindef}
\end{equation}
and 
\begin{equation}
{\cal L}_{s,mp}(z) = - \im k \iint G_m^*({\bf a},z) N_s({\bf a}-{\bf a}',z) G_p({\bf a}',z)\ {{\rm d}^2 a}\ {{\rm d}^2 a'} .
\label{dispdef}
\end{equation}
The imaginary number $\im$ is included in the defnition of ${\cal L}_{s,mp}(z)$, but not in  ${\cal P}_{mp}(z)$, in anticipation of the form of the final expression. From Eqs.~(\ref{kindef}) and (\ref{dispdef}), we respectively note that ${\cal P}_{mp}$ is hermitian and that ${\cal L}_{s,mp}$ is anti-hermitian,
\begin{eqnarray}
{\cal P}_{mp}^{\dag} = & {\cal P}^*_{pm} = {\cal P}_{mp} \label{kinher} \\
{\cal L}_{s,mp}^{\dag} = & {\cal L}^*_{s,pm} = -{\cal L}_{s,mp} \label{dispher} ,
\end{eqnarray}
following from the fact that $N_s^*({\bf a},z_0) = N_s(-{\bf a},z_0)$, because $\tilde{n}({\bf x})$ is a real-valued function.

The hermitian adjoint operator for the infinitesimal propagation through turbulence is then given by
\begin{equation}
\langle q|dU_s^{\dag}|n\rangle = \delta_{qn} - \im dz\ {\cal P}_{qn}^{\dag} + dz\ {\cal L}_{s,qn}^{\dag} .
\label{infucdef}
\end{equation}

\subsection{Hamiltonian}

For a Hamiltonian based approach, one can determine the required `Hamiltonian' directly from the definition of the unitary operator in Eq.~(\ref{infudef}), by assuming that $z$ can be treated as `time' and that we work in units where $\hbar=1$. This Hamiltonian is given by
\begin{equation}
{\cal H}_{mp} = {\cal P}_{mp} - \im {\cal L}_{s,mp} ,
\label{hamil}
\end{equation}
where ${\cal P}_{mp}$ is interpreted as the kinetic term and ${\cal L}_{s,mp}$ as the potential or `interaction' term. In the present case, however, (and this is very important) the field with which the optical field $g$ `interacts' is not a quantum field, but rather a classical field that represents the fluctuating refractive index $\tilde{n}$. Moreover, $\tilde{n}$ is not a dynamical field --- it does not have a kinetic term. As a result the potential is equivalent to a mass term and the Hamiltonian has no interaction terms. If $\tilde{n}$ were a dynamical quantum field, a consistent formulation of this process would have required quantum field theory \cite{qedequiv}. A consequence of this observation is that, when the Markov approximation is introduced in the derivation of the Lindblad equation, it differs from the Markov approximation usually associated with a system coupled to a thermal bath. Here there is no coupling and the thermal bath is replaced by the classical refractive index fluctuations. As a result the Markov approximation used here is equivalent to that which is used in classical scintillation theory \cite{scintbook}.

\subsection{Second order in $N_s$}

Substituting Eqs.~(\ref{infudef}) and (\ref{infucdef}) into Eq.~(\ref{infqelsom}), and expanding the result to first order in $dz$, we obtain
\begin{eqnarray}
\rho_{mn}(z_0 + dz) = & \rho_{mn}(z_0) + \im dz \left[ {\cal P}, \rho(z_0) \right]_{mn} \nonumber \\
& + dz \sum_s^S {1\over S}\ \left[ {\cal L}_{s,mp}(z_0) \rho_{pn}(z_0) + \rho_{mq}(z_0) {\cal L}_{s,qn}^{\dag}(z_0) \right] ,
\label{lb1}
\end{eqnarray} 
where we used Eq.~(\ref{kinher}) to express the kinetic part as a commutator. We also use the summation convention for all indices (except $s$), which means that there is an implied summation for all repeated indices. In this form the dissipative term [last term in Eq.~(\ref{lb1})] would vanish if one performs ensemble averaging, because the ensemble average of the refractive index fluctuations is zero ${\cal E}\{ N_s\}=0$. One needs an expression that is second order in $N_s$ to obtain nonzero dissipative terms.

Considering the equation in Eq.~(\ref{lb1}) for one instance of the refractive index, we turn it into a first order differential equation with respect to $z$
\begin{equation}
\partial_z \rho_{mn}(z) = \im \left[ {\cal P}, \rho(z) \right]_{mn} + {\cal L}_{s,mp}(z) \rho_{pn}(z) + \rho_{mq}(z) {\cal L}_{s,qn}^{\dag}(z) .
\label{lb2}
\end{equation} 
Next, we integrate Eq.~(\ref{lb2}) from $z_0$ to $z$ to obtain
\begin{eqnarray}
\rho_{mn}(z) = & \rho_{mn}(z_0) + \im \int_{z_0}^z \left[ {\cal P}, \rho(z_1) \right]_{mn}\ {\rm d}z_1 \nonumber \\
& + \int_{z_0}^z \left[ {\cal L}_{s,mp}(z_1) \rho_{pn}(z_1) + \rho_{mq}(z_1) {\cal L}_{s,qn}^{\dag}(z_1) \right]\ {\rm d}z_1 .
\label{lb3}
\end{eqnarray} 
Then we substitute Eq.~(\ref{lb3}) back into itself repeatedly until we have second order terms in $N_s$ containing density matrix elements that only depend on $z_0$. 

The expressions for ${\cal P}$ and ${\cal L}$ given in Eqs.~(\ref{kindef}) and (\ref{dispdef}) contain the momentum space wave functions $G_m$ and $G_m^*$, evaluated at $z$. However, under the paraxial approximation all modal functions are slow varying in $z$. As a result, for short enough distances, $G_m({\bf a},z) \approx G_m({\bf a},z_0)$. Hence, one can replace all these functions with their equivalents evaluated at $z_0$.

One can then evaluate the $z$ integral for the kinetic term and discard terms beyond the first order in $(z-z_0)$. (Eventually we'll set $z-z_0=dz$.) We also discard terms that are first order in $N_s$, because ${\cal E}\{ N_s\}=0$. The resulting equation is
\begin{eqnarray}
\fl \rho_{mn}(z) = & \rho_{mn}(z_0) + \im (z-z_0) \left[ {\cal P}, \rho(z_0) \right]_{mn} \nonumber \\
& + \int_{z_0}^z \int_{z_0}^{z_1} \left[ - {\cal L}_{s,mp}^{\dag}(z_1) {\cal L}_{s,pq}(z_2) \rho_{qn}(z_0)+ {\cal L}_{s,mp}(z_1) \rho_{pq}(z_0) {\cal L}_{s,qn}^{\dag}(z_2) \right. \nonumber \\
& \left. + {\cal L}_{s,mp}(z_2) \rho_{pq}(z_0) {\cal L}_{s,qn}^{\dag}(z_1)- \rho_{mp}(z_0) {\cal L}_{s,pq}^{\dag}(z_2) {\cal L}_{s,qn}(z_1) \right] {\rm d}z_2\ {\rm d}z_1 ,
\label{lb4}
\end{eqnarray} 
where we used Eq.~(\ref{dispher}) to change some of the ${\cal L}$s into ${\cal L}^{\dag}$s and visa versa. Note that the two $z$-integrations in the dissipative term are coupled. Moreover, the result after the two integrations over $z$ must be linear in $(z-z_0)$ in order to give a first order differential equation in $z$.

The kinetic term in Eq.~(\ref{lb4}) already has the correct form as required for the Lindblad equation. Its sign is determined by the fact that the evolution is in terms of space and not time. One needs to remember that the `Hamiltonian' in this case is actually a momentum operator. For this reason we use ${\cal P}$, instead of ${\cal H}$, to represent its free part.

\subsection{Markov approximation}

The dissipative term in Eq.~(\ref{lb4}) needs more careful attention. After replacing all momentum space wave functions in Eq.~(\ref{dispdef}) with their equivalents evaluated at $z_0$, the only remaining $z_1$- and $z_2$-dependences are associated with the $N_s$s. Moreover, the ensemble averaging only operates on the products of $N_s$s.

The dissipative term, therefore, contains integrals of the form
\begin{equation}
\Gamma_0({\bf a}_1,{\bf a}_2) = \int_{z_0}^{z} \int_{z_0}^{z_1} {\cal E}\left\{ N({\bf a}_1,z_1) N^*({\bf a}_2,z_2) \right\} \ {\rm d} z_2\ {\rm d} z_1 .
\label{verw1}
\end{equation}
One way to define $N_s$, which is consistent with Eqs.~(\ref{wiener}) and (\ref{structf}), is
\begin{equation}
N_s({\bf a},z) = \int \exp(-\im k_z z) \tilde{\chi}_s({\bf k}) \left[ {\Phi_0({\bf k})\over\Delta^3} \right]^{1/2} {{\rm d} k_z\over 2\pi} ,
\label{spek2d}
\end{equation}
where $\tilde{\chi}_s({\bf k})$ is a three-dimensional, frequency domain, normally distributed, random complex function and $\Delta$ is an associated correlation length in the frequency domain. The fact that $\tilde{n}$ is a real-valued function implies that $\tilde{\chi}^*({\bf k})=\tilde{\chi}(-{\bf k})$. Furthermore, it is assumed that this random function is delta-correlated,
\begin{equation}
{\cal E}\left\{ \tilde{\chi}({\bf k}_1) \tilde{\chi}^*({\bf k}_2) \right\} = \left( 2\pi\Delta \right)^3 \delta_3 ({\bf k}_1-{\bf k}_2) .
\label{verwrand}
\end{equation}

We now substituting Eq.~(\ref{spek2d}) into Eq.~(\ref{verw1}) and evaluate the ensemble average with the aid of Eq.~(\ref{verwrand}). Then we evaluate one of the $k_z$-integrals to obtain 
\begin{eqnarray}
\Gamma_0({\bf a}_1,{\bf a}_2) = & \delta_2 ({\bf a}_1-{\bf a}_2) \int \int_{z_0}^{z} \int_{z_0}^{z_1} \Phi_0({\bf k}_1) \nonumber \\
& \times \exp[\im k_{1z} (z_2-z_1)]\ {\rm d} z_2\ {\rm d} z_1\ {{\rm d} k_{1z} \over 2\pi} .
\end{eqnarray}
Evaluating the two $z$-integrals and dropping terms that are anti-symmetric in $k_{1z}$, we obtain
\begin{equation}
\Gamma_0({\bf a}_1,{\bf a}_2) = \delta_2 ({\bf a}_1-{\bf a}_2) \int \Phi_0({\bf k}_1) {1-\cos[k_{1z} (z-z_0)] \over k_{1z}^2} \ {{\rm d} k_{1z} \over 2\pi} .
\end{equation}
The $z$-dependent function is sharply peaked at $k_{1z}=0$ provided that $(z-z_0)$ is larger than the correlation length of the refractive index fluctuations. Under this condition one can substitute $k_{1z}=0$ in $\Phi_0({\bf k}_1)$ and evaluate the remaining integral over $k_{1z}$. The latter condition represents the Markov approximation, as used in classical scintillation theory \cite{scintbook}. Here it is applicable because $\tilde{n}$ is a classical field. The resulting expression is  
\begin{equation}
\Gamma_0({\bf a}_1,{\bf a}_2) = {dz\over 2} \delta_2 ({\bf a}_1-{\bf a}_2) \Phi_1({\bf a}_1) ,
\label{gam2}
\end{equation}
where we substituted $z=z_0+dz$ and defined $\Phi_1({\bf a}_1)=\Phi_0(2\pi{\bf a}_1,0)$. Note that, although we had two $z$-integrals, we ended up with only one factor of $dz$.

\subsection{Preliminary Lindblad form}

One can obtain the same expression for $\Gamma_0({\bf a}_1,{\bf a}_2)$ as given in Eq.~(\ref{gam2}), by using
\begin{equation}
\Gamma_0({\bf a}_1,{\bf a}_2) = {dz\over 2} {\cal E}\left\{ M({\bf a}_1) M^*({\bf a}_2) \right\} ,
\end{equation}
instead of Eq.~(\ref{verw1}), where
\begin{equation}
M_s({\bf a}) = \tilde{\xi}_s({\bf a}) {\left[ \Phi_1({\bf a}) \right]^{1/2}\over\Delta} ,
\label{spekm2d}
\end{equation}
with $\tilde{\xi}_s({\bf a})$ being a two-dimensional, frequency domain, normally distributed, random complex function, such that $\tilde{\xi}^*({\bf a})=\tilde{\xi}(-{\bf a})$ and
\begin{equation}
{\cal E}\left\{ \tilde{\xi}({\bf a}_1) \tilde{\xi}^*({\bf a}_2) \right\} = \Delta^2 \delta_2 ({\bf a}_1-{\bf a}_2) .
\label{verwrandm}
\end{equation}
In effect, the $z$-integration is implicit in $M_s$. Hence, unlike $N_s$, $M_s$ is independent of $z$.

We remove the $z$-integrals in Eq.~(\ref{lb4}) by introducing a factor of $dz/2$ and replacing the ${\cal L}$-operators by new $L$-operators, defined by 
\begin{equation}
L_{s,mp}(z_0) = - \im k \iint G_m^*({\bf a},z_0) M_s({\bf a}-{\bf a}') G_p({\bf a}',z_0)\ {{\rm d}^2 a}\ {{\rm d}^2 a'} .
\label{ndispdef}
\end{equation}
The expression in Eq.~(\ref{lb4}) then becomes 
\begin{eqnarray}
\fl \rho_{mn}(z_0+dz) = & \rho_{mn}(z_0) + \im dz \left[ {\cal P}, \rho(z_0) \right]_{mn} + {dz\over 2} \left[ 2 L_{s,mp}(z_0) \rho_{pq}(z_0) L_{s,qn}^{\dag}(z_0) \right. \nonumber \\
& \left. - L_{s,mp}^{\dag}(z_0) L_{s,pq}(z_0) \rho_{qn}(z_0) - \rho_{mp}(z_0) L_{s,pq}^{\dag}(z_0) L_{s,qn}(z_0) \right] ,
\label{lb5}
\end{eqnarray} 
where we substituted $z=z_0+dz$. One can now introduce the ensemble averaging, as expressed by the summation over all instances of the refractive index fluctuations, and convert the result into a first order differential equation in $z$. The resulting expression is equivalent to a master equation in the Lindblad form  
\begin{eqnarray}
\partial_z \rho_{mn}(z) = & \im \left[ {\cal P}, \rho(z) \right]_{mn} + {1\over 2} \sum_s^S {1\over S}\ \left[ 2 L_{s,mp}(z) \rho_{pq}(z) L_{s,qn}^{\dag}(z) \right. \nonumber \\
& \left. - L_{s,mp}^{\dag}(z) L_{s,pq}(z) \rho_{qn}(z) - \rho_{mp}(z) L_{s,pq}^{\dag}(z) L_{s,qn}(z) \right] .
\label{lb6}
\end{eqnarray}
The redefinition of the noise spectra $N_s$ in terms of purely two-dimensional functions $M_s$ allows one to show that the equation can be expressed as a Lindblad equation. As a result one can conclude that the density matrix that solves this equation obeys the requirements of unity trace and positivity. 

The expression in Eq.~(\ref{lb6}) should be regarded as a preliminary Lindblad equation, because the ensemble averaging has not been evaluated yet. As a result the Lindblad operators contain random functions. Moreover, the number of Lindblad operators is determined by the number of elements in the ensemble and the decay constants are all equal. Therefore, although the expression suffices to show the validity of the density matrix, it is not very useful for the purpose of calculations. It is necessary to evaluate the ensemble averages and find Lindblad operators that are not random functions.

We now proceed to evaluate the ensemble averages in the expression. This can be done either in terms of $N_s$ or $M_s$, since both give the same result. For convenience we proceed with $M_s$.

\subsection{Ensemble averaging}

The three ensemble averages in Eq.~(\ref{lb6}) can all be obtained from 
\begin{eqnarray}
\Lambda_{mnpq} & = \sum_s^S {1\over S}\ L_{s,mp}(z) L_{s,qn}^{\dag}(z) = {\cal E}\left\{ L_{mp}(z) L_{qn}^{\dag}(z) \right\} \nonumber \\
& = k^2 \iiiint G_m^*({\bf a}_1,z) G_p({\bf a}_2,z) G_q^*({\bf a}_3,z) G_n({\bf a}_4,z) \nonumber \\
& ~~~~~ \times  {\cal E}\left\{ M({\bf a}_1-{\bf a}_2) M({\bf a}_3-{\bf a}_4) \right\}\ {{\rm d}^2 a_1}\ {{\rm d}^2 a_2}\ {{\rm d}^2 a_3}\ {{\rm d}^2 a_4} .
\label{verw2}
\end{eqnarray}
Redefining ${\bf a}_1 \rightarrow {\bf a}_1+{\bf a}_2$ and ${\bf a}_4 \rightarrow {\bf a}_3+{\bf a}_4$, we have
\begin{eqnarray}
\Lambda_{mnpq} = & k^2 \iiiint G_m^*({\bf a}_1+{\bf a}_2,z) G_p({\bf a}_2,z) G_q^*({\bf a}_3,z) G_n({\bf a}_3+{\bf a}_4,z) \nonumber \\
& \times  {\cal E}\left\{ M({\bf a}_1) M^*({\bf a}_4) \right\}\ {{\rm d}^2 a_1}\ {{\rm d}^2 a_2}\ {{\rm d}^2 a_3}\ {{\rm d}^2 a_4} .
\label{verw3}
\end{eqnarray}
The ensemble average over the products of $M_s$s gives
\begin{equation}
{\cal E}\left\{  M({\bf a}_1) M^*({\bf a}_4) \right\} = {\cal E}\left\{  M({\bf a}_1) M(-{\bf a}_4) \right\} = \Phi_1({\bf a}_1) \delta_2 ({\bf a}_1-{\bf a}_4) .
\label{eam}
\end{equation}
We substitute Eq.~(\ref{eam}) into Eq.~(\ref{verw3}) and evaluate the integral over ${\bf a}_4$, to obtain
\begin{eqnarray}
\Lambda_{mnpq} & = k^2 \iiint G_m^*({\bf a}_1+{\bf a}_2,z) G_p({\bf a}_2,z) G_q^*({\bf a}_3,z) G_n({\bf a}_3+{\bf a}_1,z) \nonumber \\
& ~~~~~ \times \Phi_1({\bf a}_1)\ {{\rm d}^2 a_1}\ {{\rm d}^2 a_2}\ {{\rm d}^2 a_3} \nonumber \\
& = k^2 \int W_{mp}({\bf a},z) W_{qn}^{\dag}({\bf a},z) \Phi_1({\bf a})\ {{\rm d}^2 a}
\label{verw4}
\end{eqnarray}
where
\begin{equation}
W_{ab}({\bf a},z) = \int G_a^*({\bf a}'+{\bf a},z) G_b({\bf a}',z)\ {{\rm d}^2 a'} .
\label{wdef}
\end{equation}

Next we consider the case where two of the indices are contracted
\begin{eqnarray}
\sum_p \Lambda_{mnpp} = & k^2 \sum_p \iiint G_m^*({\bf a}_1+{\bf a}_2,z) G_p({\bf a}_2,z) G_p^*({\bf a}_3,z) \nonumber \\
& \times G_n({\bf a}_3+{\bf a}_1,z) \Phi_1({\bf a}_1)\ {{\rm d}^2 a_1}\ {{\rm d}^2 a_2}\ {{\rm d}^2 a_3} .
\label{verw5}
\end{eqnarray}
The contraction leads to the completeness condition for these basis functions
\begin{equation}
\sum_p G_p({\bf a}_2,z) G_p^*({\bf a}_3,z) = \delta_2({\bf a}_2-{\bf a}_3) .
\label{volledig}
\end{equation}
Note that this completeness condition only applies in the infinite dimensional case where all basis functions are considered. Hence, the corresponding density matrix is infinite dimensional. Substituting Eq.~(\ref{volledig}) into Eq.~(\ref{verw5}) and evaluating the integral over ${\bf a}_3$, we obtain
\begin{eqnarray}
\sum_p \Lambda_{mnpp} & = k^2 \iiint G_m^*({\bf a}_1+{\bf a}_2,z) \delta_2({\bf a}_2-{\bf a}_3) G_n({\bf a}_3+{\bf a}_1,z) \nonumber \\
& ~~~~~ \times \Phi_1({\bf a}_1) {{\rm d}^2 a_1}\ {{\rm d}^2 a_2}\ {{\rm d}^2 a_3} \nonumber \\
& = k^2 \iint G_m^*({\bf a}_1+{\bf a}_2,z) G_n({\bf a}_2+{\bf a}_1,z) \nonumber \\
& ~~~~~ \times \Phi_1({\bf a}_1)\ {{\rm d}^2 a_1}\ {{\rm d}^2 a_2} .
\label{verw6}
\end{eqnarray}
Next we redefine ${\bf a}_2 \rightarrow {\bf a}_2-{\bf a}_1$. The resulting integral over ${\bf a}_2$ is an orthogonality relationship that gives a Kronecker delta
\begin{equation}
\int G_m^*({\bf a}_2,z) G_n({\bf a}_2,z)\ {{\rm d}^2 a_2} = \delta_{mn} .
\label{ortogees}
\end{equation}
Applying the redefinition and Eq.~(\ref{ortogees}) to Eq.~(\ref{verw6}), we obtain
\begin{equation}
\sum_p \Lambda_{mnpp} = \delta_{mn} k^2 \int \Phi_1({\bf a})\ {{\rm d}^2 a} = \delta_{mn} \Lambda_T .
\label{verw7}
\end{equation}

\subsection{Final Lindblad form}

From the above calculation we see that
\begin{equation}
\sum_s^S {1\over S}\ L_{s,mp}(z) L_{s,qn}^{\dag}(z) = k^2 \int W_{mp}({\bf a},z) W_{qn}^{\dag}({\bf a},z) \Phi_1({\bf a})\ {{\rm d}^2 a} .
\label{lindop}
\end{equation}
By replacing all the pairs of Lindblad operators in Eq.~(\ref{lb6}) with these integrals, one obtains a new Lindblad equation, given by 
\begin{eqnarray}
\partial_z \rho_{mn}(z) = & \im \left[ {\cal P}, \rho(z) \right]_{mn} + {k^2\over 2} \int \Phi_1({\bf a}) \left[ 2 W_{mp}({\bf a},z) \rho_{pq}(z) W_{qn}^{\dag}({\bf a},z) \right. \nonumber \\
&  - W_{mp}^{\dag}({\bf a},z) W_{pq}({\bf a},z) \rho_{qn}(z) \nonumber \\
& \left. - \rho_{mp}(z) W_{pq}^{\dag}({\bf a},z) W_{qn}({\bf a},z) \right] \ {{\rm d}^2 a} .
\label{lb8}
\end{eqnarray}
Here we see that the Lindblad operators are the correlation functions between pairs of the spectra of the basis functions. The power spectral density of the turbulence model takes on the role of the decay constants. Instead of discrete Lindblad operators, leading to a summation, we have a continuous spectrum of Lindblad operators, labeled by the two-dimensional spatial frequency ${\bf a}$. Hence, the summation is replace by a two-dimensional integral over the spatial frequencies. 

\subsection{Infinitesimal propagation equation}

We now use the expressions and definitions in Eqs.~(\ref{verw2}) to (\ref{verw7}) to express Eq.~(\ref{lb6}) as
\begin{equation}
\fl \partial_z \rho_{mn}(z) = \im {\cal P}_{mp}(z) \rho_{pn}(z) - \im \rho_{mp}(z) {\cal P}_{pn}(z) + \Lambda_{mnpq}(z) \rho_{pq}(z) - \Lambda_T \rho_{mn}(z) .
\label{ipe0}
\end{equation}
This expression is the infinitesimal propagation equation (IPE) for a single photon propagating through turbulence. It demonstrates the equivalence between the Lindblad equation in Eq.~(\ref{lb8}) and the IPE.

\subsection{Generalizations}
\label{genipe}

The single photon expression in Eq.~(\ref{ipe0}) can be generalized for two photon (bi-partite) states. The density operator becomes a density tensor contracted with the bra- and ket-basis vectors for both photons
\begin{equation}
\rho = \sum_{m,n} |m\rangle |p\rangle\ \rho_{mnpq}\  \langle n| \langle q| .
\label{rhooam}
\end{equation}
Here $|m\rangle$ and $\langle n|$ are, respectively, the ket- and bra-basis vectors for the one photon, and $|p\rangle$ and $\langle q|$ are, respectively, the ket- and bra-basis vectors for the other photon.

First, we consider the case where only one of the two photons propagates through turbulence. The other photon propagates through free-space without turbulence. The IPE for this case contains the free-space kinetic terms for both photons, but only dissipative terms for one of the photons. The resulting expression is thus given by
\begin{eqnarray}
\partial_z \rho_{mnpq} = & \im {\cal P}_{mx} \rho_{xnpq} - \im \rho_{mxpq} {\cal P}_{xn} + \im {\cal P}_{px} \rho_{mnxq} - \im \rho_{mnpx} {\cal P}_{xq} \nonumber \\
& + \Lambda_{mnxy} \rho_{xypq} - \Lambda_T \rho_{mnpq} .
\label{glb1}
\end{eqnarray}

Next we consider the case where both photons propagate through turbulence, but along different paths, so that the turbulent media are uncorrelated. As a result, the expression now contains dissipative terms for both photons
\begin{eqnarray}
\partial_z \rho_{mnpq} = & \im {\cal P}_{mx} \rho_{xnpq} - \im \rho_{mxpq} {\cal P}_{xn} + \im {\cal P}_{px} \rho_{mnxq} - \im \rho_{mnpx} {\cal P}_{xq} \nonumber \\
& + \Lambda_{mnxy} \rho_{xypq} + \Lambda_{pqxy} \rho_{mnxy} - 2 \Lambda_T \rho_{mnpq} .
\label{glb2}
\end{eqnarray}
This expression is the same as the one obtained in \cite{ipe}, (apart from a difference in notation).

Finally, we consider a case where both photons propagate through the same turbulent medium, which allows for additional correlation terms to appear \cite{ipecor}
\begin{eqnarray}
\partial_z \rho_{mnpq} = & \im {\cal P}_{mx} \rho_{xnpq} - \im \rho_{mxpq} {\cal P}_{xn} + \im {\cal P}_{px} \rho_{mnxq} - \im \rho_{mnpx} {\cal P}_{xq} \nonumber \\
& + \Lambda_{mnxy} \rho_{xypq} + \Lambda_{pqxy} \rho_{mnxy} + \Lambda_{mqxy} \rho_{xnpy} + \Lambda_{pnxy} \rho_{myxq} \nonumber \\
& - \Lambda_{xnqy} \rho_{mypx} - \Lambda_{mxyp} \rho_{ynxq} - 2 \Lambda_T \rho_{mnpq} .
\label{glbc}
\end{eqnarray}

\section{Conclusions}
\label{concl}

A master equation in the Lindblad is derived for the evolution of a photonic quantum state while propagating through turbulence. This master equation allows one to study the process of entanglement decay in the spatial degrees of freedom that the photonic state experiences in a turbulent environment. The derivation follows a multiple phase screen approach, which overcomes the requirement for weak scintillation conditions, as in the case of the single phase screen model. A preliminary Lindblad equation is obtained prior to the evaluation of the ensemble averages over all possible refractive index fluctuations. Once the ensemble averages are evaluated the final form of the Lindblad form contains integrations over the spatial frequency, instead of summations. The decay constants are given by the power spectral density of the turbulence model. It is shown that the expression of the Lindblad master equation is equivalent to the infinitesimal propagation equation, provided that the full infinite dimensional space of spatial degrees of freedom is retained. The derivation is done for a single photon and the resulting expression is generalized to cases for two photons where either one photon or both photons propagate through turbulence. The equivalence between the expression in Lindblad form and the IPE implies that the solution of the full IPE represents a valid density matrix. In practice one may need to truncate the IPE to a finite number of basis elements. In such a case the resulting density matrix is in general not trace preserving. 

As part of the derivation, we show that the single phase screen model can be expressed by an operator product expansion, which implies that it represents a valid quantum process. We discuss the conditions for weak scintillation and point out that there are situations where the quantum state may retain a nonzero entanglement well into the strong scintillation region. Under such conditions the single phase screen model would not be valid anymore.

Although the emphasis in this work is the effect of atmospheric turbulence on the quantum state that propagates through it, this effect is not the only issue that would need to be addressed in the practical implementation of a free-space quantum communication system \cite{shapiro,vasylyev,usenko}. Another important aspect is the fact that any practical communication system has a finite receiver aperture, which induces a loss of optical power and information. However, these aspects are beyond the scope of the current paper.

\ack

The author wishes to express his gratitude for discussions with Clemens Gneiting and Chahan Kropf. 

\section*{References}


\end{document}